# Network Analysis of the French Environmental Code

Romain Boulet[1], Pierre Mazzega[1], Danièle Bourcier[2]

[1] Université de Toulouse; UPS (OMP),
CNRS IRD, LMTG Toulouse
14 av. Belin 31400 Toulouse, France
{boulet, mazzega}@lmtg.obs-mip.fr

[2] CERSA CNRS,
Université de Paris 2,
10 rue Thénard, 75005 Paris, France
daniele.bourcier@cersa.cnrs.fr

**Abstract.** We perform a detailed analysis of the network constituted by the citations in a legal code, we search for hidden structures and properties. The graph associated to the Environmental code has a small-world structure and it is partitioned in several hidden communities of articles that only partially coincide with the organization of the code as given by its table of content. Several articles are also connected with a low number of articles but are intermediate between large communities. The structure of the Environmental Code is contrasting with the reference network of all the French Legal Codes that presents a rich-club of ten codes very central to the whole French legal system, but no small-world property. This comparison shows that the structural properties of the reference network associated to a legal system strongly depends on the scale and granularity of the analysis, as is the case for many complex systems.

## 1 Introduction

In recent years, the debates about the legal complexity by spontaneous orders and the possibility to control it have recovered strength and interest from many different scientific and politic communities. In particular, when it is seen as a part of liberalization, the simplification of law is expected by different governmental bodies in Europe to reduce some administrative burdens, to induce net positive economical returns or, for example, to increase the trade volumes between State members, or even worldwide. However, what is the legal complexity, how to control it, and what might be the impacts of the simplification of the Law, are questions without any element of answer. This article does not attempt to define or redefine legal complexity (exposed for instance by Hayek's works) but is aimed to better understand an aspect of this legal complexity induced by the numerous citation links between articles. In a seminal paper [1], we proposed to open a field of research aiming at building rigorous definitions of legal complexity that could be operational and exploited over large legal data bases.

Instead of directly confronting ourselves with the semantic complexity of the Law, we decided to analyze some structure associated with the legal systems. The positive fact in choosing this less ambitious objective is that we start building and using definitions of some aspects of legal complexity that, exploited with legal corpuses, provide new insight on the structural properties of some legal systems. These insights in turn open new perspectives on the Law. In particular the granularity of the legal texts we are considering is fundamental in the analysis because, as we shall see, the properties of the legal structure differ with the scale of the analysis (as is the case for many complex systems). Indeed in a previous study, we considered as the smallest object legal codes themselves and proceed to the analysis of the network formed by their cross-citations [2]. Using mathematical tools developed for the analysis of social networks, we found that several hidden stable structures are underlying the network of the French legal codes: a "rich club" is gathering the ten most cited and most citing codes that are very strongly connected to each others. Several other code communities also exist, in particular one of 12 codes related to "social matters" and "social activities", and another one of 11 codes regulating various matters linked to "territories and natural resources".

Changing the scale and granularity of the analysis, new structural properties are likely to appear. Here we choose to consider a single code and to analyze the network of citations within it. In a previous study, we have increased our resolution till the distinction of subdivisions within the articles. A statistical analysis of the distribution of the levels of organization of the Environmental Code and of the corresponding number of objects brought interesting clues about a kind of self-organization even within a code [3]. In the present study we go beyond the statistical analysis by considering with scrutiny the associated network structure.

## 2  A Network Approach to Legal Complexity

### 2.1  Networks dealing with legal complexity

Studying legal networks brings a new point of view on the issues of the complexity of Law. Approaching the complexity of the law through network analysis is a novel area of study and few analyses have been made so far.

Previous work mainly focused on the analysis of the citation network of the Supreme Court jurisprudence [4], [5], [6]. In [6], some structural elements are highlighted: the network possesses a main core of 122 vertices, the most cited cases and the most central cases are enumerated and, despite a low density, the network is locally dense. James Fowler also used the different notions of centrality to study the citation network of the Supreme Court precedents [4]. The United States Code is also considered as a network in [7], the vertices of the network are the different sections of the US Code and the links are citations links. A common aspect of these papers is the examination of the degree[1] distribution in order to bring out a scale-free effect. The network structure of the Uniform Commercial Code is analyzed in [8]; in this paper citations links and hierarchical links are considered and several tools are exposed in order to understand the shape of the network. These tools are computing indices of the graph (like small world indices or centralities) or performing a good visualization of the graph in order to detect communities.

In the next subsection we shall introduce the legal network we study throughout this paper and then in Section 3 we shall not only compute the degree distribution and central vertices but also exhibit a small-world structure and cluster the network into communities. The analyses we perform, with an emphasis on legal interpretations of our results, will give us a better understanding of the shape of this network .

---

[1] The degree of a vertex is the number of vertices linked to this vertex.

**2.2 Associating a Network to the Environmental Code**

Several networks – sets of vertices (or node) possibly linked by edges - can be associated to any legal corpus. We consider the cross-citations between various objects. Such an object can be, among others, an international Treatise or Convention, a law, a whole code, any subdivision of a code (book, title, chapter, etc.), an article, a key word, *etc.* There are numerous citations between these objects. Then it is straightforward to associate a network of citations (hereafter also called *reference network*) where the objects of the considered corpus are vertices and the citations constitute edges (or links) between them. For a network of several tens to thousands of vertices and edges, only appropriate mathematical tools allow to recover the main underlying properties and hidden structures.

In this study our corpus is the French Environmental Code (but our approach would apply as well to any other code). The nodes are the articles belonging to the code and also the various objects of the hierarchy of the code, say books, titles, chapters, sections, etc. Then we define two kinds of edges [1], [3]: influence-type edges and selection-type edges. Influence-type links are closely related to the hierarchy structure of the Code as they link objects *A* and *B* if *B* is a subdivision or a part of *A*. Being a "part of" surely indicates a non-arbitrary relationship, a dependence based on the organization of the legal substance of the corresponding texts or hierarchical levels within the code. This yields the tree structure of the Code (as can be retrieved from its table of contents).

A selection-type edge links two articles *C* and *D* if there is an explicit reference to *D* in *C*. For instance, the following extract of Article L211-3 produces a link between articles L211-3 and L211-2 and a link between articles L211-3 and L211-1: "*Article L211-3 (extract): In addition to the general regulations mentioned in Article L. 211-2, national or particular provisions with regard to certain parts of the territory are established by a Conseil d'Etat decree in order to ensure the protection of the principles set out in Article 211-1.*". This kind of link is also called *citation link* or *reference link*. Moreover any other object of the hierarchy of the code (not only articles) can be cited as for example in the Article L222-4 which explicitly cites the Article L222-1 and the Chapter III of Title III of Book I, thus creating the corresponding two selection-type links in the associated network. Note that the links to objects that do not belong to our corpus (here the Environmental Code) are discarded in the analysis.

**2.3 The French Environmental Code**

In France, an important step in the intelligibility and the accessibility of the Law was made in 1989 when the Government decided to accelerate and reinvigorate the codification process with the creation of the *Commission Supérieure de Codification*. The codification process contributes to the clarification and ordering of the Law for citizens by collecting norms and regulations on a specific field in a single book called a *Code.* The present study is realized with the version of the Environmental Code as available on the Légifrance site [9] at the reference date of March, 27$^{th}$ 2009. Though the code is regularly updated and transformed, we believe that the results presented here are robust and representative for a period covering a decade or so (this period will depend on the rate of change of the texts in the corpus, this rate being itself most dependent on the regulated matter).

The French environmental Code is divided into a legislative part and a regulatory part. The legislative part has been voted by the parliament in 2003. The regulatory part should theoretically mirror the structure of the legislative part, and its content develops more specifically the corresponding regulations to be implemented. Here we only consider the legislative part and we construct the network of references as described in the previous sub-section. The legislative part of the Environmental Code is divided into the following seven books [9]: (I) Common provisions (Art. L121-1 to L110-2); (II) Physical environments (Art. L211-1 to L220-2); (III) Natural spaces (Art. L310-1 to L300-3); (IV) Flora and fauna (Art.

L411-1 to L430-1); (V) Prevention of pollution, risks and nuisances (Art. L511-1 to L582-1); (VI) Provisions applicable in New Caledonia, French Polynesia, the Wallis and Futuna Islands, French Southern and Antarctic Territories and Mayotte (Art. L611-1 to L656-1); (VII) Protection of the environment in the Antarctic (Art. L711-1 to L713-9).

Added to these 7 books, we find 1775 other vertices (the smallest objects that we consider being articles) distributed into 31 titles, 122 chapters, 201 sections, 107 sub-sections, 26 paragraphs and 1288 articles. Among this total of 1782 vertices, 513 vertices do not share any link with another vertex. They are "isolated vertices". This is particularly the case of objects of the hierarchy which are never explicitly cited like for example the Chapter IV of Title II of Book II *"National technical measures for the prevention of atmospheric pollution and the rational use of energy"*. There are 93 such vertices that are not texts but headings of group of articles. It is also the case for articles which are never cited and do not cite any other part of the code like for example the Article L429-2 (*"The right to hunt on lands or water-covered areas is administered by the municipality for and on behalf of the owners"*).

We hereafter focus on the greatest connected component which contains 980 vertices, that is more than 80% of vertices sharing at least one link. The second greatest connected component has 40 vertices, the third one has 26 vertices and the remaining components have fewer than 5 vertices.

## 3 Analyzing the Network Structure of the Environmental Code

In the following the graph G denotes the greatest connected component of the reference network of the French Environmental Code. In order to visualize the hierarchical organization of the code on this reference network, we give a color to each vertex according to the book they belong to. A first representation of this network is done by using a spring-like force algorithm, often used to represent networks in a readable way and such as there are as few crossing edges as possible (Figure 1). However, owing to the high number of vertices, further analyses are necessary to better understand the network.

### 3.1 A Small-World Network

The small-world structure has been found in many social or social-like networks [10], [11], [12], [13]. The definition of a small world structure is based on two main properties:
- Only few steps are needed to link two any vertices;
- The probability that two vertices are linked by an edge is much higher if these vertices already have a common neighbor.

The first point can be measured by the *characteristic path length* [13], [12] which is the median of the average shortest path starting at a vertex. The second point can be measured by a *clustering coefficient* which is the mean of densities of neighborhood of vertices. The global density of the graph is the ratio between the number of edges in the graph and the maximal number of edges the graph could have if all the vertices were pair wise linked. The clustering coefficient measures the local density that is the density around a vertex. In a small-world network this local density is much higher than the global density.

In order to estimate how high or low are these indices (the characteristic path length and the clustering coefficient), we compare them to those of other real networks sharing the small-world property. We also compare these indices to those calculated on an Erdos-Renyi random graph: the global connectivity (here measured by the characteristic path length) must be almost the same and the clustering coefficient must be much higher in a small-world type network.

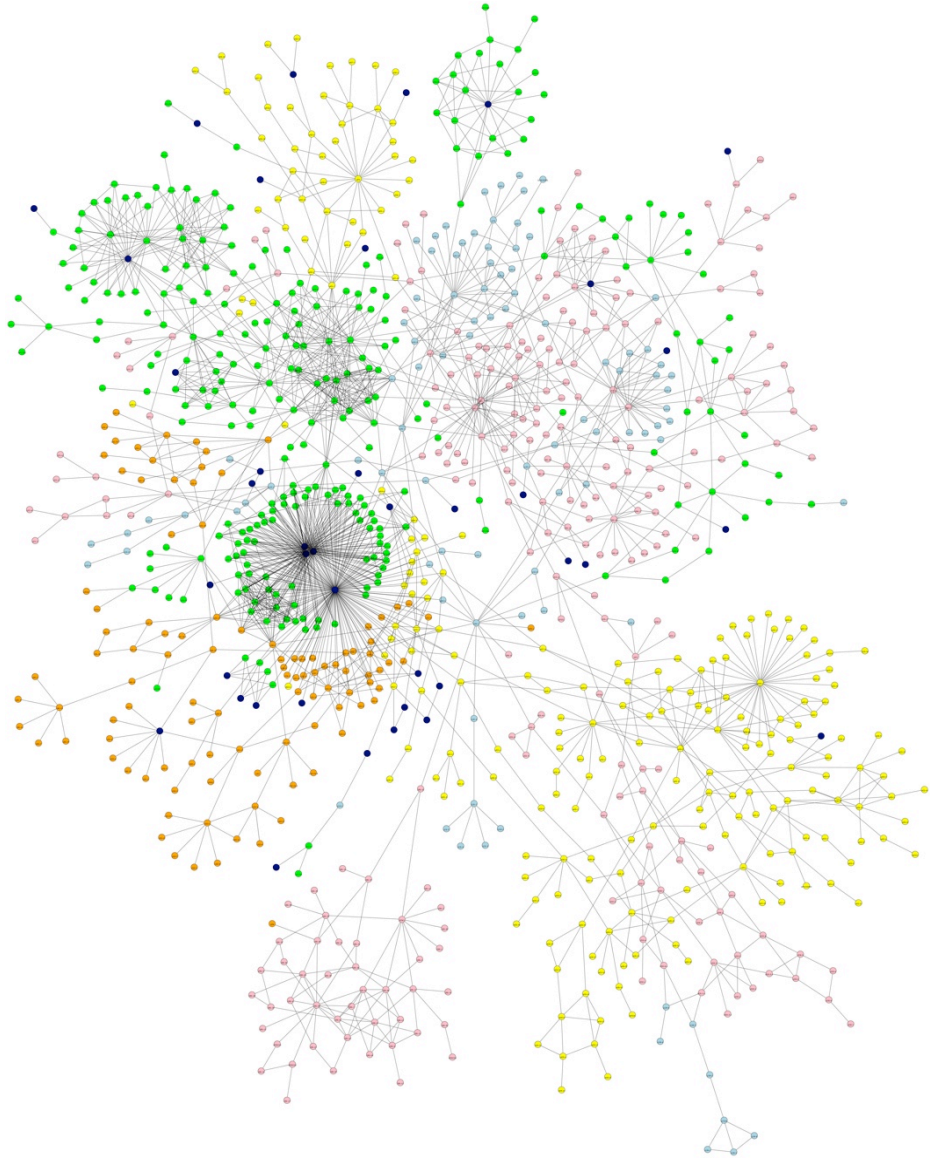

**Figure 1.** The graph G associated to the main connected component (980 vertices, 2186 edges) of the Environmental Code (legislative part only). The colors given to the vertices indicate the Book they are belonging to (hereafter only short names are given): (blue) Common provisions; (green) Physical environments; (orange) Natural space; (yellow) Flora and fauna; (pink) Prevention of pollutions, Risks nuisances; (dark blue) Provisions applicable in New Caledonia etc.; (grey) Environmental protection in Antarctica.

**Table 1.** Small-world indices of G (graph of the greatest connected component of the Environmental Code), of the associated random graph, of a medieval social network [14] and of a network between German companies in 1993-1997 [15] (two companies are linked if they have a common owner). The number of vertices is denoted by n, the number of edges by m and the density by d, (with d=2m/[n(n-1)]). The small-world properties are related to the characteristic path length (L) and the first clustering coefficient (C).

| Graph | n | m | d | L | C |
|---|---|---|---|---|---|
| G | 980 | 2186 | 0.0046 | 6.78 | 0.49 |
| Random graph | 980 | 2186 | 0.0046 | 4.61 | 0.0046 |
| Medieval | 615 | 4193 | 0,0222 | 3.71 | 0.78 |
| Companies | 291 | 1036 | 0.02 |  | 0.84 |

As we can see in Table 1, the reference network of the French environmental code is a small-world. Indeed, the global connectivity measured by the characteristic path length is close to the one obtained for a random graph and the clustering coefficient is much higher than the global density of the graph.

**3.2 Degree distribution.**

Another strong characteristic of real networks (or social-like networks) are scale-free networks or networks having a power-law degree distribution. That means that the empirical probability[2] for a node to have a degree k, that is to be linked to k other nodes, is proportional to a power of k. As the network considered in this paper is undirected we only examine the degree distribution for this graph. This distribution is represented on Figure 2a and is clearly a long tail distribution: there are a lot of vertices with a low degree and few vertices with a high degree revealing the presence of hubs. Figure 2b represents the same distribution but drawn on a log-log scale in order to check if this distribution is a power-law[3]. We recall that a power-law distribution can be obtained by a preferential attachment model [16]: when a new vertex enters the network, it links to other vertices with a probability proportional to their degree. In other words this new node will preferentially link to a high degree node (this can be seen as the *winner-takes-all* effect or *rich get richer*).

Here, we observe that the decreasing of the distribution is stronger for high degrees, this is a common phenomena which occurs for instance in the distribution of inward and outward citations of the judicial precedent network [4]. This deviation from power-law can be explained by the fact that the network elements have a finite capacity to add links, which limits their maximum degree [17].

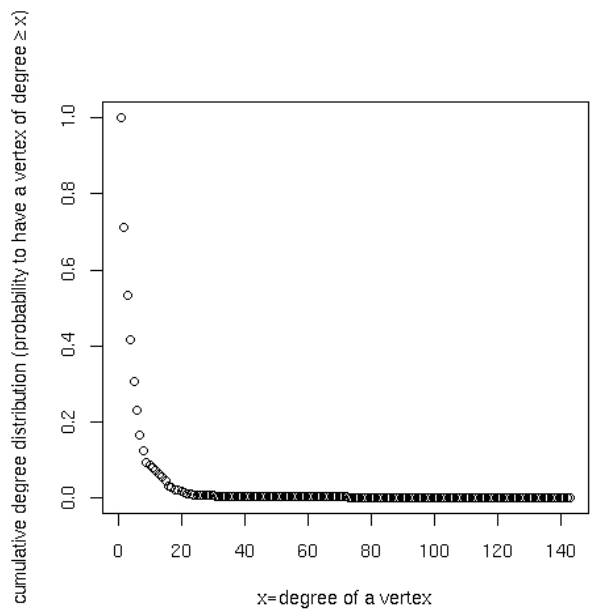

**Figure 2a:** Cumulative degree distribution of the citation network of the French environmental code. The y-axis represents the probability to have a vertex of degree greater than or equal to x.

---

[2] For practical reasons we consider the cumulative distribution, that is the probability for a node to have a degree greater than or equal to k. This allows a smoothing of the distribution. Note that a distribution is power-law if and only if its cumulative distribution is power-law.

[3] A power-law drawn in a log-log scale is represented by a straight line.

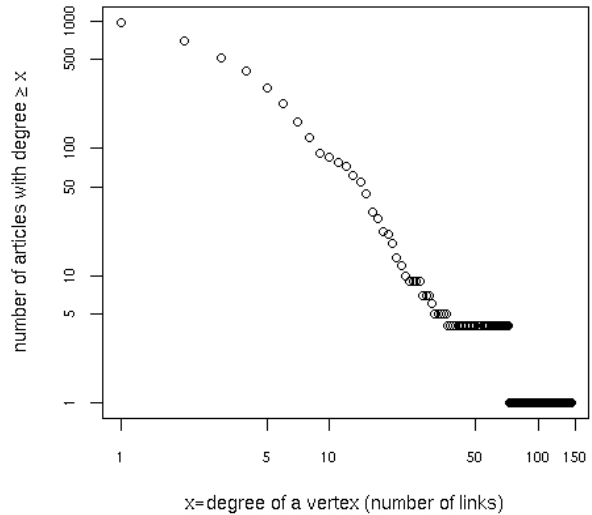

**Figure 2b:** Cumulative degree distribution of the citation network of the French environmental code plotted on a log-log scale. The y-axis represents the probability to have a vertex of degree greater than or equal to x.

### 3.3 Central Articles

We then examine whether G possesses a "rich-club" that is a central and influent community consisting of vertices (e.g. articles) with highest degrees and highly interconnected. A rich-club provides a network structure mainly organized around this influential group. The degrees of the nine articles with highest degrees are listed in Table 2. It appears that this distribution is quite heterogeneous with a strong decrease and a gap between the largest degree and the second largest degree. In the network of citations of the French Environmental Code, it appears that there is no edge between the eight vertices with highest degree. Therefore they cannot constitute a highly interconnected group. Oppositely to the network of French Legal Codes [2] the network of citations in the French Environmental Code does not have a rich-club of articles (or of any other objects in the code). From a legal point of view, it means that every code is written in an autonomous way following an internal logic but that the global project of codification follows an emergent and successive organization.

**Table 2** The nine vertices with highest degrees.

| Rank | Article | Degree | Rank | Article | Degree | Rank | Article | Degree |
|---|---|---|---|---|---|---|---|---|
| 1 | L640-1 | 143 | 4 | L632-1 | 72 | 7 | L216-5 | 30 |
| 2 | L612-1 | 72 | 5 | L429-1 | 36 | 8 | L213-11 | 27 |
| 3 | L622-1 | 72 | 6 | L216-3 | 31 | 9 | L652-1 | 27 |

Another important centrality notion is betweenness [18] which measures how important a vertex is for the network connectivity by counting the number of shortest paths going through this vertex. Figure 3a distinguishes some articles with a particularly high betweenness measure and Figure 3b shows some vertices with a low degree but a high betweenness centrality. We select the eight vertices with highest betweenness indices, the corresponding articles are listed in Table 3.

Among other things, Article L640-1 enumerates the references of a large list of articles of the Books I, II, III and IV of the Environmental Code that apply to the French Southern and Antarctic Territories. This content explains why it has both the highest degree and the highest betweenness (see Table 2 and 3, and Fig.2a). The Title I of Book V has a relatively high degree that results from being often cited. All together the Articles L511-1 to L517-2 form a whole description of the "*Classified facilities for the protection of the environment*" under this heading that is cited in preference to the individual articles. But these facilities for the protection of the environment are of concern for many other parts of the Environmental Code that otherwise would not be related. This is the reason for the relatively high betweenness centrality of this heading (Title I of Book V).

**Table 3.** The eight articles with the highest indices of betweenness centrality.

| Rank | Article | Betweenness | Degree | Rank | Article | Betweenness | Degree |
|---|---|---|---|---|---|---|---|
| 1 | L640-1 | $2.57 \times 10^5$ | 143 | 5 | L218-44 | $5.80 \times 10^4$ | 14 |
| 2 | L141-1 | $1.28 \times 10^5$ | 15 | 6 | L142-2 | $5.49 \times 10^4$ | 7 |
| 3 | L424-3 | $7.31 \times 10^4$ | 14 | 7 | Book V, Title I | $4.50 \times 10^4$ | 23 |
| 4 | L413-4 | $7.22 \times 10^4$ | 4 | 8 | L581-32 | $4.38 \times 10^4$ | 3 |

Article L142-2 is a good example of a central vertex with a low degree (*d=7*). Indeed it cites only three other articles and is cited by four other articles but these articles belong to four different books. As a result Article L142-2 is linking different communities of articles. As is seen in Fig.2b, many other articles in the Environmental Code have a low degree and a relatively high betweenness. By linking different communities they strongly structure the overall architecture of the code. But how do we reveal these communities? Do they exactly coincide with books, titles, or chapters of the Environmental Code or are they forming hidden structures within the Code?

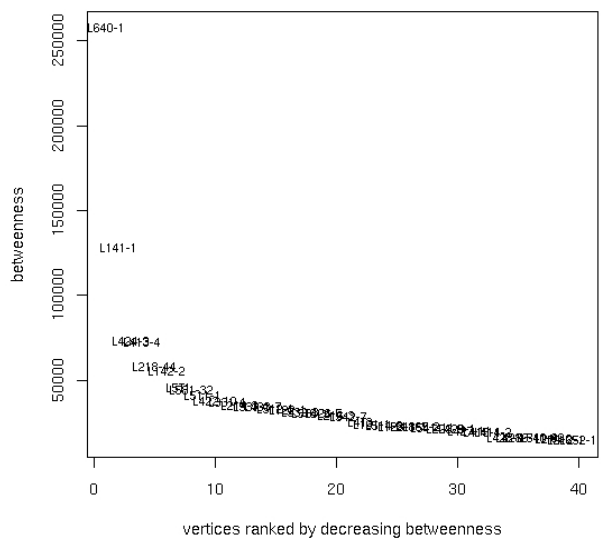

**Figure 3a:** Betweenness centrality for the 40 vertices with highest betweenness measure.

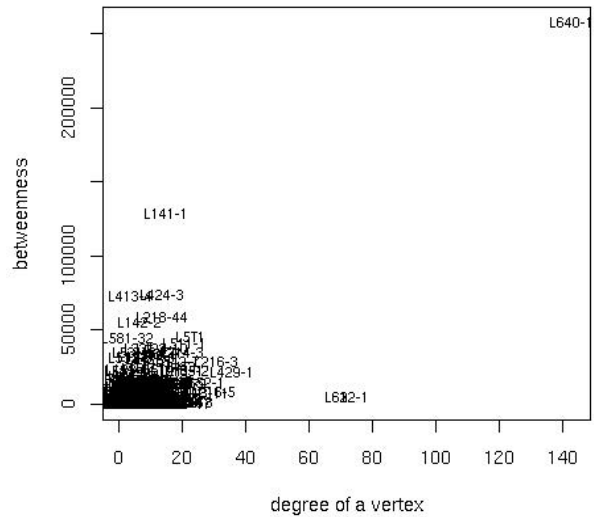

**Figure 3b.**: Betweenness centrality of a vertex (y-axis) as a function of its degree (x-axis).

### 3.4 Gathering Articles into Communities

Like social networks, the vertices of many real networks can be gathered into groups densely wired internally and with few connections between these groups. By analogy with social networks, such groups are called communities. In graph theory, the process of graph partitioning aims at dividing the graph into parts such that the number of links between the parts is as small as possible, and as large as possible within each part.

A preliminary step in the partitioning process is to remove the eight central vertices found in Section 3.3. Excluding these vertices will afford a clearer representation: let us assume that a central vertex links two communities A and B. If we do not remove the central vertex, there is a risk for A and B to be aggregated together in a single community at the end of the partitioning algorithm though they are distinct. Moreover this approach will allow us to emphasize the role played by these central vertices (e.g. articles).

Without entering into technical details (see the mathematical references if required), we use a spectral partitioning algorithm based on the eigenvectors of the normalized graph Laplacian [19], [20]. We obtain a set of communities of articles. Some of them are directly linked by edges between articles belonging to these communities. By adding the central vertices we previously deleted we can visualize how these different groups of vertices are connected. Figure 4 gives a visualization of this structure.

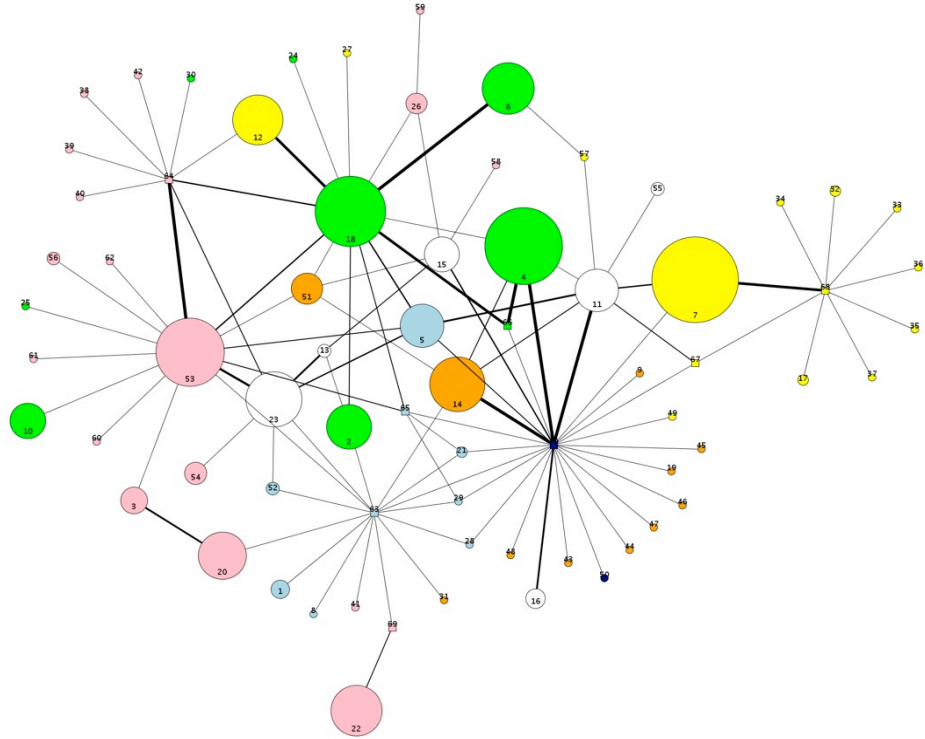

**Figure 4.** Communities found in the largest connected part of the French Environmental Code and obtained by a spectral partitioning. A disk represents a community of vertices. Its area is proportional to the number of vertices therein. A community is colored if more than 75% of the vertices belong to the same Book; otherwise the community is white. The colors are the same than in Fig 1. The thickness of a link is related to the number of citations between the communities.

Three kinds of communities arise:
- Communities consisting of articles from the same section, chapter or book (note that for instance belonging to a same chapter is stronger than belonging to a same book). For example, the community with number 1 is composed of articles from the Chapter I of Title III of Book I *"Institutions acting in the domain of environmental protection"*;
- Communities mostly composed of articles from a same part of the Code. For instance vertices of community number 6 belong to the Chapter III of Title I of Book II excepted the articles L652-1and L652-8;
- More heterogeneous communities. Among the 22 vertices of community 15, 10 of them belong to Chapter II of Title II of Book I and 9 belong to Chapter I of Title VII of Book V.

In Table 4 we give the list of the main communities found in the graph G. The biggest one counts 135 articles, almost all belonging to the Book IV (*Flora and fauna*) of the Environmental Code. This group of articles has been gathered together in the code, and functions like a closed, "autonomous" community where most of references is pointing towards an article of the same group (and book). This community appears as the largest yellow disk on the right side of Fig.3.

**Table 3.** Size of the communities with the percentage of vertices belonging to the Book the most represented in that community (that Book and its associated color in Figure 4 are given in brackets). We do not mention the 34 communities of size 1.

| Size | % of art. Book, color | Size | % of art. Book, color | Size | % of art. Book, color | Size | % of art. Book, color |
|---|---|---|---|---|---|---|---|
| 135 | 99.3 (IV, yellow) | 47 | 97.9 (V, pink) | 22 | 50 (I, white) | 3 | 66.7 (I, white) |
| 109 | 77.1 (II, green) | 46 | 87.0 (IV) | 18 | 100 (III, orange) | 3 | 66.7 (II, white) |

| 90 | 85.6 (II, green) | 41 | 80.5 (V, pink) | 13 | 92.3 (V, pink) | 3 | 100 (I, blue) |
| 84 | 91.7 (V, pink) | 36 | 94.4 (II, green) | 9 | 88.9 (V, pink) | 3 | 100 (V, pink) |
| 56 | 64.3 (V, white) | 34 | 94.1 (II, green) | 8 | 100 (V, pink) | 2 | 100 (I, blue) |
| 55 | 96.4 (III, orange) | 33 | 63.6 (IV, white) | 7 | 57.1 (II, white) | 2 | 100 (IV, yellow) |
| 48 | 95.9 (II, green) | 23 | 95.7 (II, green) | 6 | 100 (I, blue) | 2 | 100 (IV, yellow) |

About 77% of the articles of the second largest community (109 articles) are belonging to Book II. The nearly 23% of the remaining articles are from one or several other books of the Environmental Code. An interesting exercise would be to analyze the content of these articles and see if they could be explicitly gathered under a common heading. It could be also extended to the other communities that we find and that present a non negligible number of articles external to the principal book of membership. We also notice that five important communities mainly issued from Book V appear in Fig.3 (green disks; see also Tab.4), though this Book counts only two titles, entitled "*Water and aquatic environments*" and "*Air and the atmosphere*" respectively. This features shows that the communities retrieved in the reference network are not simply reproducing the organization of the content of the code, but constitute hidden structures underlying it.

We also find several communities of size 1. Indeed, when we remove the central vertices, the resulting graph contains some isolated vertices, that is vertices sharing no links; these vertices cannot be gathered into a larger community and therefore constitute a community of size one.

Most of the revealed largest communities are mainly constituted of articles from the book treating of the physical environments (green disks in Fig.3), from the book on the prevention of pollutions, risks nuisances (pink) and from the book on the flora and fauna (yellow). The matters treated in these communities are presented in several articles with a relatively important number of cross-citations, like in a local sub-network. In these quite complex matters (see for example the way the water and aquatic environments are regulated in the Environmental Code) there is now easy or desirable way to organize the articles in a tree-like hierarchy. The contents of the articles are intrinsically interdependent. On the opposite we found no community gathering a majority of articles related to the provisions applicable in New Caledonia etc. (Book VI; dark blue vertices in Fig.1) or to the environmental protection in Antarctica (Book VII; grey vertices in Fig.1).

In Fig.3 we also see several small communities (sometimes in fact a single article) that are the center for a star-like shape (see e.g. the right most part of Fig.3), but the articles linked to this central node are not citing each others. Then they do not form a community.

## 4 Discussion

We can point out the particular position in the network of Book VI. A quick examination of figure 1 reveals that this book (the articles of which are colored in deep blue) is very disconnected and its articles are scattered and spread out in the network and they are linked to articles belonging to the five first books. As mentioned in the previous section, Figure 4 substantiates this fact as there is no community colored in deep blue. This observation confirms a remark made by the *Commission Supérieure de Codification* in paragraph of [21] devoted to French overseas territories "*The Commission met in a very classic way, questions relating to the codification of provisions relating to overseas territories. The complexity of the law of the overseas presents for each code some issues.*"

The low density, the small world structure and the absence of a rich-club in the citation network of the French Environmental Code clearly contrast with the architecture found in the analysis of the reference network at the scale of codes [2]. Changing the granularity of the analysis modifies the nature of network structure and its associated properties. The density variation is not unexpected because while increasing the granularity (that is grouping vertices), the number of vertices decreases and the probability of existence of a citation between the two any bigger vertex increases. If this structure modification may be evident for the density (and therefore the small world effect), the disappearance of the rich-club is remarkable. This observed relationship between change of structure and granularity could be further investigated by considering other codes and other legal systems.

The mathematical analysis of the graph associated to the Environmental Code allows to identify hidden communities of articles interrelated by cross-citations. An analysis of the content of the articles so grouped should bring new understanding on the way the matter is regulated. This question can be approached by lawyers or by the members of the *Commission Supérieure de Codification* (but is not the subject of this study), probably giving new insights on the way the legal substance is conceived and organized. Such analysis should be performed considering the largest communities with the higher percentage of "external" articles (articles in the community but belonging to different books or titles) as identified in Table 3.

## 5 Conclusion

This study provides a detailed analysis of the network of citations of a legal code. As a result we made a step forward in understanding the aspect of law complexity induced by intra-references of the norms in the Environmental Code. We focus our analysis on the largest connected graph found in this code that counts 980 vertices and 2186 edges (citations). This network is characterized by a small-world structure with several central vertices connecting communities. The structure into books of the Code is globally consistent with the exception of Book VI (*Provisions applicable in New Caledonia, French Polynesia, the Wallis and Futuna Islands, French Southern and Antarctic Territories and Mayotte*) which is not well and densely connected and the articles of which are linked to very different parts of the Environmental Code.

The structure of the Environmental Code is contrasting with the network formed by the citations between the French legal Codes (each code being a vertex). In this dense network we found a central group of ten most cited and most citing codes, strongly interlinked, and forming a rich-club, but no small-world structure. This comparison reinforces the idea that the underlying structures and properties of legal networks are dependent on the scale and granularity of the performed analysis.

The present study is a necessary stage to future applications. One of them is to compare different environmental codes for example the National Environmental Code and the two recent Provincial Environmental Codes of New Caledonia (the environmental matter belongs to the territories), or to compare this environmental code with other codes related to other legal fields (like the rural Code). These comparisons would allow us to state new hypotheses about the specificity of the organization of laws related to environmental matters or, if the occasion arises, to make assumptions concerning the constant shape of citations networks of legal codes. Another application would be to perform such analysis at different dates; this would give a way to measure the dynamical evolution of Law by measuring the evolution of small world indices, degree distributions and by examining the behavior of communities (which communities becomes denser or which communities are created).

**Acknowledgements.** R. Boulet benefits from a post doctoral grant of the *Institut National des Sciences de l'Univers* (CNRS, Paris). This study is funded by the RTRA *Sciences et Techniques de l'Aéronautique et de l'Espace* (http://www.fondation-stae.net/) in Toulouse (MAELIA project - http://www.iaai-maelia.eu/ ). The yEd Graph editor has been used for producing the Fig.1 and Fig.3. Statistical properties of networks have been computed with *R* and the library *igraph* (http://www.r-project.org/). We thank the anonymous reviewer for bringing to our attention the reference [8].